\documentclass[12pt]{article}

\usepackage{amssymb,amsthm}
\usepackage{named}
\usepackage{graphicx}
\usepackage{amsmath}
\usepackage[ruled]{algorithm2e}
\usepackage{newicktree}
\usepackage{setspace}
\newtheorem{thm}{Theorem}
\newtheorem{lemma}[thm]{Lemma}

\newtheorem{cor}[thm]{Corollary}

\newcommand{\ignore}[1]{}
\newcommand{\half}{\frac{1}{2}}

\begin{document}

\title{Neighbor Joining with Phylogenetic Diversity Estimates\\ \vskip 0.2in \small{Research article submission for the Journal of Molecular Biology and Evolution}}

\author{Dan Levy\thanks{Department of Mathematics, UC Berkeley} \and Ruriko Yoshida\thanks{Department of Mathematics, Duke University} \and Lior Pachter\footnotemark[1] \thanks{Corresponding author, {\tt lpachter@math.berkeley.edu} \newline \indent \indent 970 Evans Hall, Berkeley, CA 94720 \newline \indent  \indent Tel: (510) 642-2028, Fax: (510) 642-8204}}

\maketitle

Running head: Neighbor joining with PD estimates
\newpage \begin{abstract}
The {\em Neighbor-Joining algorithm} is a recursive procedure for
reconstructing trees that is based on a transformation of pairwise
distances between leaves. We present a generalization of the
neighbor-joining transformation, which uses estimates of phylogenetic 
diversity rather than pairwise distances in the tree.  This
leads to an improved neighbor-joining algorithm whose total running
time is still polynomial in the number of taxa.  On simulated data,
the method outperforms other distance-based methods.

We have implemented neighbor-joining for subtree weights in a program
called {\tt MJOIN} which is freely available under the Gnu Public
License at \newline {\tt http://bio.math.berkeley.edu/mjoin/}
\end{abstract}

\newpage
\section{Introduction}

Distance based methods for phylogenetic reconstruction are based on the observation that edge weighted phylogenetic $X$-trees (trees that have a set $X$ as their leaves, all interior vertices of degree at least three and non-negative weights
$w_T:E(T) \rightarrow {\mathbb R}_{>0}$ on every edge) can be encoded by certain metrics on $X$.\

\begin{thm}[Four-point condition \cite{Buneman1971}]
\label{fourpoint}
Given a\\metric $D:X \times X \rightarrow {\mathbb R}$ there exists an edge weighted phylogenetic $X$-tree $T$ such that $D(i,j)=\sum_{e \in E(T )} w_T(e)$ iff \[D(i,j) + D(k,l) \leq max(D(i,k)+D(j,l),D(j,k)+D(i,l)) \] for every four leaves $i,j,k,l$. Furthermore, $T$ is unique.
\end{thm}
Such metrics are called {\em tree metrics} and many methods have been proposed for projecting dissimilarity maps (functions $D:X \times X \rightarrow {\mathbb R}$ with $D(x,x)=0$ and $D(x,y)=D(y,x)$) to ``nearby'' tree metrics.  The neighbor-joining algorithm, introduced by \cite{Saitou1987}, is the most popular and widely used. It is particularly convenient for reconstructing phylogenetic trees when the size of $X$ is large, in which case methods that require an exhaustive exploration of the space of trees are computationally prohibitive.

There are four parts to the neighbor-joining algorithm (see algorithm 1):
\begin{enumerate}
 \item A procedure for estimating pairwise distances between elements of $X$.
 \item A criterion for identifying neighboring pendant edges (cherries) in a tree.
 \item A recursive reduction.
 \item A branch length estimation formula.
\end{enumerate}

The cherry picking criterion is based on the following theorem:
\begin{thm}[\cite{Saitou1987,Studier1988}]
\label{simplecherry}
If $D$ is a tree metric and
\[ Q_D(i,j)= (n-2) D(i,j) - \sum_{k \neq i} D(i,k) -\sum_{k \neq j} D(j,k) \]
then the pair $x,y$ that minimizes $Q_D(x,y)$ is a cherry in the tree.
\end{thm}
Although the exact formula for $Q$ may seem a bit mysterious at first, it is a very natural criterion. For example, the neighbor-joining algorithm which is based on it is {\em consistent} (i.e. if $D$ is a tree metric then the algorithm returns the tree), the input order of the taxa does not change the outcome of the algorithm, and the criterion is a linear function of the distances. \cite{Bryant2005} has recently shown that the neighbor-joining selection criterion $Q(i,j)$ is the {\em only} one satisfying the properties above. Furthermore, \cite{Gascuel1997b} has shown that the neighbor-joining criterion can be interpreted as greedily minimizing a balanced minimum evolution criterion which provides added understanding as to why it has been a very successful method.

The recursive reduction step and branch length estimation formula have been examined extensively and have resulted in a number of improvements to the basic neighbor-joining algorithm. For example, the reduction step has been extensively investigated and has been shown to be optimal when variances on the estimates are unknown, yet improvable when variance information is incorporated \cite{Gascuel1994, Gascuel1997, Gascuel1997b}.

\begin{algorithm}
\label{neighborjoining}
   \SetLine
   \AlgData{A set $X$ together with sequences corresponding to the elements of $X$}
   \AlgResult{Edge weighted phylogenetic $X$-tree $T$}
   \For{$i,j \in {X \choose 2}$}{
     Compute the maximum likelihood distance $D(i,j)$ between taxa $i$ and $j$\;
   }
   \While{$|X|>2$}{
     \For{$i,j \in {X \choose 2}$}{
       Set $Q_D(i,j)=(|X|-2)D(i,j)-\sum_{k \in X \setminus \{j\}}D(i,k)  - \sum_{k \in X \setminus \{i\}}D(kj)$.
     }
     Choose a pair $x,y \in X$ that minimizes $Q_D(x,y)$\;
     Add a new element $z_{|X|}$ to the set $X$ and remove $x$ and $y$\;
     Let $u_{|X|}=x$ and $v_{|X|}=y$.\;
     Set $D(i,z_{|X|})=\frac{1}{2}(D(i,x)+D(i,y)-D(x,y))$\;
   }
   \While{$|X| \leq n-2$}{
     Set $D(u_{|X|},z_{|X|})=\frac{1}{|X|-2}\sum_{k \neq u_{|X|},v_{|X|}} D(u_{|X|},k)+D(u_{|X|},v_{|X|})-D(k,v_{|X|})$\;
     Set $D(v_{|X|},z_{|X|})=\frac{1}{|X|-2}\sum_{k \neq u_{|X|},v_{|X|}} D(v_{|X|},k)+D(u_{|X|},v_{|X|})-D(k,u_{|X|})$\;
     Add $u_{|X|}$ and $v_{|X|}$ into $X$.
   }
   \caption{Neighbor-joining algorithm}
 \end{algorithm}

Nevertheless, the main problem with neighbor-joining scheme is that in the first step, the distances are estimated from noisy data and the resulting dissimilarity map is therefore very unlikely to be a tree metric. For biological sequences, the pairwise distance estimates are typically based on a probabilistic model of evolution such as the \cite{Jukes1969} model:  given two sequences of length $L$ with $k$ differences between them, the distance is estimated as
\[ D_{JC}=-\frac{3}{4}ln \left( 1-\frac{4}{3}p \right) \]
where $p=\frac{k}{L}$. The variance is given by
\[ Var(D_{JC}) \approx \frac{p(1-p)}{L(1-\frac{4}{3}p)^2}. \]
Notice that as $p \rightarrow \frac{3}{4}$ the variance approaches infinity, which reflects the fact that long branch lengths are difficult to resolve with finite sequences. This phenomenon exists whenever branch lengths  are estimated using Markov models of evolution. Although the neighbor-joining algorithm is consistent, the fact that dissimilarity maps estimated from data are not tree metrics means that there is no guarantee that the algorithm produces the correct tree.

A number of attempts have been made to understand the good results obtained with the neighbor-joining algorithm, especially given the problems with the inference procedures used for estimating pairwise distances. One of the main results is the following:
\begin{thm}[\cite{Atteson1999}]
\label{Atteson}
Neighbor-joining has $l_{\infty}$ radius $\frac{1}{2}$.
\end{thm}
This means that if the distance estimates are at most half the minimal edge length of the tree away from their true value then the neighbor-joining algorithm will reconstruct the correct tree. However, as we will see in section 4, this criteria is rarely attained even in cases where neighbor-joining has a high success rate.

Despite the unavailability of precise criteria for judging the success of neighbor-joining, there have been efforts aimed at improving the distance estimates which form the input to the algorithm. For example, the TRIPLEML method \cite{Ranwez2002} improves on the pairwise distance estimates by adjusting them using additional taxa:  for each pair of leaves, a third leaf is selected and an approximate (numerical) maximum likelihood estimate for the branch lengths of the three leaf subtree is computed from which the pairwise distance of the original leaves is estimated.  In the WEIGHBOR algorithm \cite{Bruno2000}, the neighbor-joining criterion is replaced so as to weight long branch lengths. These methods, and others similar to them, have the drawback that either their performance remains limited by the inherent uncertainty in pairwise distance estimates, or  else the simple, natural, and mathematically justified structure of the neighbor-joining algorithm is abandoned.

It was suggested in \cite{Pachter2004} that an alternative encoding of edge weighted phylogenetic $X$-trees may be used to improve phylogenetic reconstruction while preserving many of the properties of distance based methods. Let $X^m$ denote the $m$th Cartesian product of $X$ and ${X \choose m}$ all the $m$ element subsets of $X$. For a phylogenetic $X$-tree $T$ with $R \subset X$ let $[R]$ denote the smallest subtree of $T$ spanning $R$.
\begin{thm}[\cite{Pachter2004}]
\label{Pachter-Speyer}
Let $T$ be a phylogenetic $X$-tree ($|X|=n$) and $m \geq 2$ be an integer.  Let $n \geq 2m-1$, and let $D:X^m \rightarrow {\mathbb R}$ be the map $R \mapsto \sum_{e \in [R]} w_T(e)$  for each $R \in {X \choose m}$.  Then $T$ is determined by the set of values $D(R)$ (and this is not true if $2m-2=n>2$).
\end{thm}
Instead of reconstructing trees from dissimilarity maps ($m=2$), it was suggested that maximum likelihood methods could be used to more accurately estimate the phylogenetic diversity values $D(R)$ \cite{Faith1992} for every $R \subset X, |R|=m$. The phylogenetic diversity values are also conveniently called the {\em $m$-subtree weight} values. Such estimates result in ${n \choose m}$ values which form an {\em $m$-dissimilarity map}, i.e. a function $D:X^m \rightarrow {\mathbb R}$ with $D(x,x,\ldots,x)=0$ and $D(x_1,\ldots,x_m)=D(x_{i_1},\ldots,x_{i_m})$ for any permutation $(i_1,\ldots,i_m) \in S_m$. The problem is then to develop consistent tree reconstruction algorithms that find a tree whose $m$-subtree weights are ``close'' to the $m$-dissimilarity map.

In this paper we propose a practical, efficient method for tree reconstruction based on $m$-dissimilarity maps.  We begin by refining theorem \ref{Pachter-Speyer} and show that even if $n<2m-1$ partial information about the tree is recoverable.  We then describe a neighbor-joining algorithm whose cherry picking criterion makes use of $m$-subtree weights. The algorithm is a generalization of standard neighbor-joining (in the special case $m=2$ the formulas in the algorithm simplify to neighbor-joining).  It also satisfies many of the same properties: the method is consistent, the input order of the taxa does not change the outcome, and the cherry picking criterion is a linear function of the distances. In section 4 we argue that it is more accurate than neighbor-joining, and the fact that it is polynomial in the number of taxa means that it is practical for the same kinds of large problems for which neighbor-joining is used.  In fact, the running time for $m=3$ is $O(n^3)$, the same as for standard neighbor-joining (only with a higher time constant for the initial estimation of the weights).

Our main results depends on yet another encoding of phylogenetic $X$-trees.  Given four leaves $i,j,k,l$ in a phylogenetic $X$-tree, we use the notation
\[ |(i,j;k,l)| := |E([\{i,j\}] \cap  [\{k,l\}])|.\]
We say that $(i,j;k,l)$ is a tree quartet if $|(i,j;k,l)|=\emptyset$. If $q(T)$ denotes the set of tree quartets then there is a partial order $\leq$ on all $X$-trees where $T' \leq T$ iff $q(T') \subseteq q(T)$.

\begin{thm}[\cite{Buneman1971,Semple2003}]
\label{quartet}
Let $T$ and $T'$ be two phylogenetic $X$-trees. Then $q(T)=q(T')$ iff $T \cong T'$.
\end{thm}

\section{Tree metrics from m-weights}

Our main results about $m$-subtree weights are based on a mapping that associates to any $m$-dissimilarity map a $2$-dissimilarity map which, for $m$-subtree weights from a tree, preserves a certain subforest.  This subforest is characterized by containing those edges whose removal results in sufficiently small components in the tree.  Specifically, for a tree $T$, the removal of any edges results in two components, and we denote by $T_{\leq k}$ the subforest of $T$ whose edge set consists of edges whose removal results in one of the components having size at most $k$. For example $T_{\leq 1}$ consists of all the pendant edges (adjacent to leaves), and $T_{\leq k}=T$ for any $k>\frac{n-1}{2}$ because the removal of any edge in a tree leaves a component of size at most $\frac{n-1}{2}$.  For the tree $T$ in figure \ref{fig:subforest} with 24 leaves, $T = T_{\leq 12}$.

\begin{figure}
\begin{center}
\end{center}
\caption{A tree $T$ and four subforests.} \label{fig:subforest}
\end{figure}

\begin{thm}
\label{reductionthm}
Let $D$ be an $m$-dissimilarity map on a set $X$ of size $n$ and define
\begin{equation}
S_D(i,j) = \sum_{Y \in {X \setminus \{i , j\} \choose m-2}}D(i,j,Y).
\end{equation}
If $D(R)=\sum_{e \in [R]}w_T(e)$ for every $R \in {X \choose m}$ in some edge weighted phylogenetic $X$-tree $T$, then $S_D$ is a tree metric.  Furthermore, if $T'$ is the tree corresponding to this tree metric, then $T' \leq T$ with $T'_{\leq n-m} \cong T_{\leq n-m}$ and there is an invertible linear map between the edge weights in $T_{\leq n-m}$ and the corresponding edge weights in $T'_{\leq n-m}$ (with the exception that in the case that $T \neq T_{\leq n-m}$, the pendant edge weights are not uniquely determined.).
\end{thm}

For a fixed tree $T$ and integer $m$, let $S = S_D$ where $D$ is the $m$-dissimilarity map induced by $T$. Observe that for an edge weighted phylogenetic $X$-tree, $T$, any linear combination of the $m$-subtree weights is a linear combination of the edge weights $w_T(e)$ in the tree.  For a linear function on the $m$-subtree weights $F: \mathbb{R}^{n \choose m} \to \mathbb{R}$, let $v_F(e)$ denote the coefficient of $w_T(e)$ in $F$.  For instance, $v_{S(i,j)}(e)$ denotes the coefficient of $w_T(e)$ in $S(i,j)$.  Note that $v_{F+G}(e) = v_F(e) + v_G(e)$.  We will also use the notation $L_i(e)$ to denote the set of leaves in the component of $T-e$ that contains leaf $i$ and $P_{ab}$ is the path from vertex $a$ to $b$.

\begin{lemma}
\label{mainlem}
Given a pair of leaves $a,b$ and any edge $e$ we have \[ v_{S(a,b)}(e)= \left\{
\begin{array}{ll}
{n - 2 \choose m-2} & e \in P_{ab}; \\
& \\
{n-2 \choose m-2} - {|L_a(e)|-2 \choose m-2} & e \notin P_{ab}.
\end{array}
\right. \]
\end{lemma}

{\bf Proof}: If $e$ is on the path from $a$ to $b$, then it will be included in all the subtrees $[a,b,Y]$.  If $e$ is not on the the path from $a$ to  $b$, then the only way it will be excluded is if all the other leaves fall on the $a$ side of $e$ (which is the same as the $b$ side).  That is, if $Y \subset L_a(e) \setminus \{a,b\}$.  There are ${|L_a(e)| - 2 \choose m-2}$ such sets.
\qed

\begin{lemma}
Given a quartet $(a_1,a_2;a_3,a_4)$ in $T$ with interior vertices $b_1$ and $b_2$ (figure 1), then,
\begin{eqnarray*}
v_{S(a_1,a_2)+S(a_3,a_4)}(e) & = & \left\{
\begin{array}{ll}
2{n-2 \choose m-2} - {n-|L_{a_i}(e)|-2 \choose m-2}
      \qquad  e \in P_{a_ib_{\lceil i/2 \rceil}}; & \\
      & \\
2{n-2 \choose m-2} - {|L_{a_1}(e)|-2 \choose m-2} - {|L_{a_3}(e)|-2 \choose m-2}
      \quad  e \in P_{b_1b_2}; & \\
      & \\
2{n-2 \choose m-2} - 2{|L_{a_1}(e)|-2 \choose m-2}
\qquad       e \notin [a_1,a_2,a_3,a_4]. &
 \end{array}
 \right. \\ & & \\
v_{S(a_1,a_3) + S(a_2,a_4)}(e) & = &  \left\{
\begin{array}{ll}
2{n-2 \choose m-2} - {n-|L_{a_i}(e)|-2 \choose m-2}
      & e \in P_{a_ib_{\lceil i/2 \rceil}}; \\ \\
2{n-2 \choose m-2}
      & e \in P_{b_1b_2}; \\ \\
2{n-2 \choose m-2} - 2{|L_{a_1}(e)|-2 \choose m-2}
      & e \notin [a_1,a_2,a_3,a_4].
\end{array}
\right.
\end{eqnarray*}
and
\[ v_{S(a_1,a_4) + S(a_2,a_3)} = v_{S(a_1,a_3) + S(a_2,a_4)} \]
\end{lemma}

\begin{figure}[ht]
\begin{center}
\end{center}
\caption{A quartet ($a_1,a_2;a_3,a_4$)}
\label{fig:cases}
\end{figure}

{\bf Proof}:  We use the fact that $v_{S(a_1,a_2) + S(a_3,a_4)} = v_{S(a_1,a_2)} + v_{S(a_3,a_4)}$ and apply the previous lemma.  We also note that for $e \notin [\{a_1,a_2,a_3,a_4\}], L_{a_1}(e) = L_{a_i}(e)$ for all $i$.
\qed

\begin{cor}
\label{maincor}
For a quartet $(a_1,a_2;a_3,a_4)$, we define
$$ S(a_1,a_2;a_3,a_4) = S(a_1,a_2) + S(a_3,a_4) - S(a_1,a_3) - S(a_2,a_4). $$
Then,
\begin{eqnarray*}
v_{S(a_1,a_2;a_3,a_4)}(e) = \left\{
\begin{array}{ll}
- {|L_{a_1}(e)|-2 \choose m-2} - {n-|L_{a_1}(e)|-2 \choose m-2}
       & e \in P_{b_1b_2}; \\
       & \\
0
       & \mbox{otherwise}.
\end{array}
\right.
\end{eqnarray*}
\end{cor}

Corollary \ref{maincor} implies that $S$ satisfies the four-point condition (\ref{fourpoint}), although it may be that $v_{S(a_1a_2;a_3a_4)}(e)=0$  which means that there are interior edges in $T'$ which have been collapsed (with length equal to $0$).  Suppose, however, that $(a_1a_2;a_3a_4) \in q(T)$ and $[\{a_1,a_2,a_3,a_4\}]$ is in a connected component of $T_{\leq n-m}$ (in other words the subtree spanning the quartet consists of edges whose removal leaves a small component). This means that if $e \in P_{b_1b_2}$ then either $L_{a_1}(e) \geq m$ or
$n - L_{a_1}(e) \geq m$ and so $S(a_1,a_2;a_3,a_4) < 0$
which means that $(a_1,a_2;a_3,a_4) \in q(T')$.
Therefore $q(T') \subset q(T)$ and it follows from theorem \ref{quartet} that $T'_{\leq n-m} \cong T_{\leq n-m}$.

It remains to show that there is an invertible linear map between the edge weights in the forests $T_{\leq n-m}$ and $T'_{\leq n-m}$:
\begin{lemma}
\label{internaledgeslem}
If $e$ is an internal edge of $T_{\leq n-m}$ with $e'$ the corresponding edge in $T'$ then
$$w_{T'}(e') = \frac{1}{2}\left( {|L_a(e)|-2 \choose m-2} + {|L_c(e)|-2 \choose m-2}\right) w_T(e)$$
where $a$ is a leaf in one component of $T-e$ and $c$ a leaf in the other.
\end{lemma}
{\bf Proof}:  Since $e$ is an internal edge, we may choose $a, b, c$ and $d$ such that $e$ is the only edge on the splitting path of $(a,b;c,d)$ (figure \ref{fig:internaledge}).  Then
\begin{eqnarray*}
w_{T'}(e') & = & \frac{1}{2} S(a,b;c,d) \\
           & = & \frac{1}{2} \left( {|L_a(e)|-2 \choose m-2} + {|L_c(e)|-2 \choose m-2} \right)w_T(e).
\end{eqnarray*}
\qed

\begin{figure}
\begin{center}
\end{center}
\caption{The quartet $(a,b; c,d)$ has only the one edge $e$ on its splitting path.} \label{fig:internaledge}
\end{figure}

\begin{cor}
\label{internaledges}
\[ w_{T}(e) = \frac{2w_{T'}(e')} {\left( {|L_a(e)|-2 \choose m-2} + {|L_c(e)|-2 \choose m-2}\right) }  \]
\end{cor}
which is well defined if $e \in T_{\leq n-m}$.

\begin{lemma}
\label{pendantedges}
Denote the edges adjacent to the leaves by $e_1,\ldots,e_n$ (with corresponding edges in $T'$ $e_1',\ldots,e_n'$) and the set of internal (non-pendant) edges by $int(E(T))$.  Let
\[ C_i=\sum_{e \in int(E(T))} \left( {n-2 \choose m-2} - {|L_i(e)|-2 \choose m-2} \right) w_{T}(e)\]
and let ${\bf A}$ be the matrix $2{n-3 \choose m-2}{\bf I} + {n-3 \choose m-3}{\bf J}$.  Then
\[ \left( \begin{array}{c} w_{T'}(e_1')\\
\vdots\\
w_{T'}(e_n') \end{array} \right)  =
\half {\bf A}
\left( \begin{array}{c}
w_{T}(e_1)\\
\vdots\\
w_{T}(e_n) \end{array} \right) +
\half \left( \begin{array}{c}
C_1\\
\vdots\\
C_n \end{array} \right)
 \ \]
\end{lemma}

\begin{figure}
\begin{center}
\end{center}
\caption{The leaf edge $e_i$ is incident on two other edges.  We may choose leaves $a$ and $b$ such that $P_{ia} \cap P_{ib} = e_i$.} \label{fig:leafedge}
\end{figure}

{\bf Proof}:  The interior vertex of an edge $e$ also adjacent to a leaf $i$ is incident to two other edges.  Choose a leaf $a$ such that $P_{ia}$ intersects one of the edges, and $b$ such that $P_{ib}$ intersects the other (figure \ref{fig:leafedge}).  Then
\[ w_{T'}(e') = \frac{1}{2}\left( S(i,a) + S(i,b) - S(a,b) \right) \]
which after some algebra gives the above lemma.
\qed

\begin{cor}
\label{leafedges}
\[ \left( \begin{array}{c}
w_{T}(e_1)\\
\vdots\\
w_{T}(e_n) \end{array} \right)  = {\bf A}^{-1}
\left( \begin{array}{c}
2w_{T'}(e_1')-C_1\\
\vdots\\
2w_{T'}(e_n')-C_n \end{array} \right) \]
where ${\bf A}^{-1} = \frac{1}{2 {n-3 \choose m-2}}\left( {\bf I} - \frac{m-2}{(m-1)(n-2)} {\bf J} \right)$.
\end{cor}

In order to recover $w_T(e)$ for every edge, we start by calculating the interior edge weights, after which we can calculate the values $C_i$. The matrix ${\bf A}$ is always invertible if $m \leq n-1$; however, calculating $C_i$ requires that $int(E(T)) = int(E(T'))$.  If $n < 2m - 1$, then while we can determine all the interior edge weights of $T_{\leq n-m}$ from $T'$, it is possible that some interior edges of $T$ have been collapsed in $T'$:  in particular, the set of edges in $E(T) \setminus E(T_{\leq n-m})$.  If $E(T) \setminus E(T_{\leq n-m}) \neq \emptyset$, then $T_{\leq n-m}$ is composed of at least two connected components and every connected component has strictly fewer than $m$ leaves.  As a result, every $m$-subtree weight will include at least one undetermined edge, and so there is no way to uniquely determine the weights of the pendant edges.

\section{Neighbor-joining with subtree weights}

Theorem \ref{reductionthm} forms the basis of the neighbor-joining algorithm with subtree weights. First, we need a generalization of the neighbor-joining criterion:
\begin{thm}[Cherry Picking Theorem]
\label{cherrypickingthm}
Let $T$ be an edge weighted phylogenetic $X$-tree with $|X|=n$ let $m$  be an integer satisfying $2 \leq m \leq n-1$. Let $D: X^m \rightarrow \mathbb{R}_{>  0}$ be the m-dissimilarity map corresponding to the weights of the subtrees of size $m$ in $T$.  If $Q_D(x,y)$ is a minimal element of the matrix \
$$Q_D(i,j) = \left( \frac{n-2}{m-1} \right) \sum_{ Y \in {X \setminus \{i , j\} \choose m-2}} D(i,j,Y) -
           \sum_{Y \in {X \setminus \{i\} \choose m-1}} D(i,Y) -\sum_{Y \in {X \setminus \{j\} \choose m-1}} D(j,Y)$$
then $x,y$ is a cherry in the tree $T$.
\end{thm}
Note that when $m=2$ this is exactly the neighbor-joining criterion ($Q$-criterion of theorem \ref{simplecherry}) as described by \cite{Studier1988}.

{\bf Proof}: Let $S(i,j)=\sum_{Y \in {X \setminus \{i , j\} \choose m-2}}D(i,j,Y)$.  By theorem \ref{reductionthm} we know that $S$ is a tree metric.  Observe that
\begin{eqnarray*}
Q_D(i,j) & = & \frac{n-2}{m-1}S(i,j) -  \sum_{Y \in {X \setminus \{i\} \choose m-1}} D(i,Y) -
               \sum_{Y \in {X \setminus \{j\} \choose m-1}} D(j,Y) )\\
         & = &  \frac{1}{m-1}((n-2)S(i,j) - \sum_{k} \sum_{Y \in {X \setminus \{i,k\} \choose m-2}} D(i,k,Y)\\ 
         & &      - \sum_{k} \sum_{Y \in {X \setminus \{j , k\} \choose m-2}} D(j,k,Y)) \\
         & = &  \frac{1}{m-1}((n-2)S(i,j)-\sum_{k}S(i,k)\\ & &-\sum_{k}S(j,k))\\
         & = & \frac{1}{m-1} Q_S(i,j)
\end{eqnarray*}
In other words, $Q_D(i,j)$ is just a scalar multiple of the neighbor-joining criterion for the tree metric $S$.  By theorem \ref{simplecherry} ($m=2$) we know that the minimal element of $Q_S(i,j)$ is a cherry in $T'$ (the tree corresponding to the tree metric $S$).  Since $m \leq n-1$, we know that $T'_{\leq 1}$ is isomorphic to $T_{\leq 1}$  and therefore the minimal element of $Q_D(i,j)$ is a cherry.
\qed

It follows from theorem \ref{reductionthm} that if $m \leq \frac{n+1}{2}$ then the neighbor-joining algorithm applied directly to $S$ is {\em topologically consistent}, i.e. will reconstruct the correct tree topology starting with the weights of all subtrees of size $m$.  The fact that there is an invertible linear map between for the edge weights, means that we can reconstruct $T$, thus leading to a consistent neighbor joining algorithm with subtree weights (algorithm 2).

\begin{algorithm}
\label{subtreealg}
   \SetLine
   \AlgData{A set $X$ together with sequences corresponding to the elements of $X$}
   \AlgResult{Edge weighted phylogenetic $X$-tree $T$}
   \For{$R \in {X \choose m}$}{
    Estimate $D(R)$ using a (numerical) maximum likelihood method\;
   }
   \For{$i,j \in {X \choose 2}$}{
     Set $S(i,j)=\sum_{Y \in {X \setminus  \{i , j\} \choose m-2}}D(i,j,Y)$\;
   }
   Apply algorithm 1 (neighbor-joining) to the ``distances'' $S(i,j)$ resulting in tree $T'$;
   Set $T=T'$\;
   Set $w_{T}(e) = \frac{2w_{T'}(e')}{\left( {|L_a(e)|-2 \choose m-2}
 +  {|L_c(e)|-2 \choose m-2}\right) }$\;
  \For{$1 \leq i \leq n$}{
   Set $C_i=\sum_{e \in int(E(T))} \left( {n-2 \choose m-2} - {|L_i(e)|-2 \choose m-2} \right) w_{T}(e)$\;
   }
   Set $\left( \begin{array}{c}
 w_{T}(e_1)\\
 \vdots\\
 w_{T}(e_n) \end{array} \right)  =  \frac{1}{2 {n-3 \choose m-2}}\left(  {\bf I} - \frac{m-2}{(m-1)(n-2)} {\bf J} \right)
 \left( \begin{array}{c}
 2w_{T'}(e_1')-C_1\\
 \vdots\\
 2w_{T'}(e_n')-C_n \end{array} \right)$\;

\caption{Neighbor-joining algorithm with subtree weights}
\end{algorithm}

The running time for computing the weights of the subtrees is $O(Ln^m)$ where $l$ is the length of the alignment and the computation of $S(i,j)$ is $O(n^{m})$ (both steps are trivially parallelizable). The subsequent neighbor-joining is $O(n^3)$ and edge weight reconstruction is $O(n^2)$. It is interesting  to note that for fixed $L$ the running time of the algorithm is $O(n^3)$ for both $m=2$ and $m=3$.

\section{Results}

We have implemented the neighbor-joining algorithm for subtree weights in a program called {\tt MJOIN}. The implementation incorporates the {\tt fastDNAml} \cite{Olsen1994} program for computing the subtree weights, and allows the user to select the sizes of the subtrees to be used.

\begin{figure}[ht]
\begin{center}
\end{center}
\caption{T1 and T2 trees of Ota and Li.}
\label{fig:T1T2}
\end{figure}

We tested {\tt MJOIN} with simulated data on the two parameter family of trees described by \cite{Ota2000}.  These are trees for which neighbor-joining has difficulty in resolving the correct topology.  We simulated 1000 data sets on each of the two tree shapes, T$_1$ and T$_2$ (Figures 2, 3) at the three edge length ratios, a/b = 0.01/0.07, 0.02/0.19, and 0.03/0.42.  This was repeated twice for sequences of length 500 and 1000BP. We also repeated the runs with the Kimura 2-parameter model and obtained similar results (not shown).

Table 1 notes the success rate of {\tt MJOIN} for $m$=2, 3, and 4 (denoted by NJ$^{(m)}$) for each data set and compares these results to the success rate of other tree reconstruction methods. It is clear from the table that as $m$ increases, the success rate of {\tt MJOIN} increases.  Hence, for $m > 2$, {\tt MJOIN} consistently out-performs neighbor-joining (NJ$^{(2)}$). For the T1 tree, NJ$^{(4)}$ out-performs even fastDNAml.

Figure 4 shows the standard deviation in the $m$-weights. We believe it is the relative improvement in the $m$-weight errors that is contributing to the improved performance of {\tt MJOIN} as $m$ increases. Checking the $l_{\infty}$ distance of the 2-distance maps from the true tree metric, we find that even in cases where neighbor-joining has a high success rate, the number of distance maps that satisfy Atteson's condition is fewer than 1\%. This suggests that the success of neighbor joining is due to other favorable features of the projection, and we believe that a deeper understanding of neighbor joining is necessary in order to rigorously understand the reasons for the improvements with $m$-subtree weights.

 \begin{table*}
\footnotesize{
 \begin{center}
 \begin{tabular}{|c c|} \hline
 \begin{tabular}{c c c}
 Tree & length (bp) & a/b\\
 \hline
 T1 & 500 & 0.01/0.07\\
  &   & 0.02/0.19\\
  &   & 0.03/0.42\\
 &1000 & 0.01/0.07\\
 &    & 0.02/0.19\\
 &    & 0.03/0.42\\
 \hline
 T2 & 500 & 0.01/0.07\\
  &   & 0.02/0.19\\
  &   & 0.03/0.42\\
 &1000 & 0.01/0.07\\
 &    & 0.02/0.19\\
 &    & 0.03/0.42\\
 \end{tabular}
 & \begin{tabular}{c c c c c c c c c}
 NJ$^{(2)}$ & NJ$^{(3)}$ & NJ$^{(4)}$ & BN & WE & NM & QP & FM\\
 \hline
 0.69 & 0.76 & 0.82 & 0.73 & 0.72 & 0.80 & 0.80 & 0.78\\
 0.53 & 0.58 & 0.73 & 0.52 & 0.47 & 0.64 & 0.70 & 0.66\\
 0.11 & 0.12 & 0.23 & 0.14 & 0.13 & 0.16 & 0.29 & 0.11\\
 0.94 & 0.96 & 0.98 & 0.96 & 0.92 & 0.97 & 0.94 & 0.97\\
 0.87 & 0.90 & 0.96 & 0.87 & 0.83 & 0.92 & 0.92 & 0.90\\
 0.33 & 0.35 & 0.52 & 0.35 & 0.29 & 0.38 & 0.53 & 0.27\\
 \hline
 0.82 & 0.84 & 0.85 & 0.86 & 0.88 & 0.93 & 0.86 & 0.90\\
 0.69 & 0.72 & 0.74 & 0.81 & 0.89 & 0.95 & 0.85 & 0.90\\
 0.19 & 0.29 & 0.36 & 0.46 & 0.70 & --   & 0.47 & 0.59\\
 0.96 & 0.97 & 0.98 & 0.98 & 0.98 & 1    & 0.97 & 0.99\\
 0.89 & 0.92 & 0.93 & 0.99 & 0.99 & 1    & 0.96 & 0.99\\
 0.40 & 0.48 & 0.57 & 0.75 & 0.92 & 0.97 & 0.70 & 0.90\\
  \end{tabular}
 \\
 \hline
 \end{tabular}
 \end{center}}
 \caption{Simulations with the Jukes-Cantor model.  NJ$^{(m)}$ = MJOIN with subtree size $m$;
   BN = BioNJ; WB = Weighbor; NM = NJML (NM); QP = the quartet puzzling algorithm; FM = fastDNAml.}
 \end{table*}

 \begin{figure*}[ht]
   \begin{center}
   \end{center}
   \caption{Standard Deviation as a percent of total weight.  For the Jukes-Cantor method,
     sequence length of 500BP, m=2,3,4 and subtrees drawn from T$_1$ and T$_2$. }
   \label{fig:Mw8}
 \end{figure*}

\newpage
 \section{Discussion}

Theorem \ref{reductionthm} establishes that pairwise distance based reconstruction methods can be used to reconstruct trees from $m$-subtree weights. This immediately suggests a number of potential improvements to the algorithm we have described. For example, by taking into account the variances of the $S(i,j)$, it should be possible to improve on the neighbor-joining algorithm for subtree weights with better agglomeration (as is done in BIONJ).

In tests we performed with $n=10$ taxa and $m=5$ (results not reported) we observed a deterioration in the accuracy of the tree reconstruction algorithm, which we attribute to inaccuracies in the subtree weights estimated with fastDNAml. In fact, tests with fastDNaml on five taxa revealed that the algorithm fails to even reconstruct the correct tree topology a significant fraction of the time. Thus, we believe that until further improvements are made in ML estimation of trees, the best subtree weight size to use will be $m=4$. We are encouraged by various efforts in this direction \cite{Contois2005,Hosten2005}.

We have found subtree weight reconstruction to be practical and efficient for much larger examples than described here.  We have run the algorithm with $m=3$ on trees of up to 50 taxa on a standard PC, and it is worth noting that for larger problems it is trivial to parallelize the $m$-weight estimation. Thus, we believe that our method is practical and recommended for large tree constructions that currently rely on either a pairwise distance method, or a heuristic maximum likelihood search. Since the latter can fail with regularity on trees with only five taxa, it is unlikely to be accurate for large trees.

Our investigations have opened up a number of interesting questions. For example, it would be useful to obtain an analog of the four point condition that characterizes the space of $m$-dissimilarity maps arising from trees. It would also be of interest to develop a subtree-weight analog of the Neighbor-Net algorithm \cite{Bryant2004}.

Finally, we point out that our results can be viewed as providing approximations to maximum-likelihood tree reconstruction by refining distance-based methods. We believe that a deeper understanding of $m$-dissimilarity maps should yield further results in this direction.

 \section{Acknowledgments}

 This work was partially funded by an NIH grant (R01HG2362). L.P. was also supported by a Sloan foundation fellowship.  D.L. was also supported by NIH grant (GM68423).

\newpage
\pagestyle{plain}
\bibliographystyle{named}
\bibliography{mJOIN}

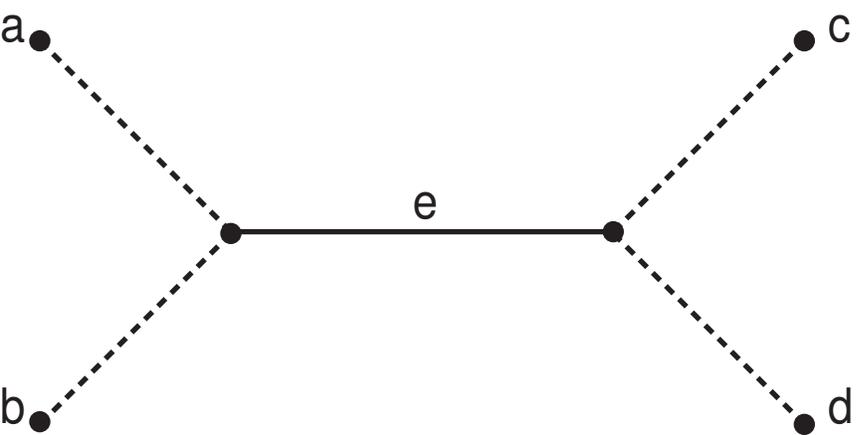
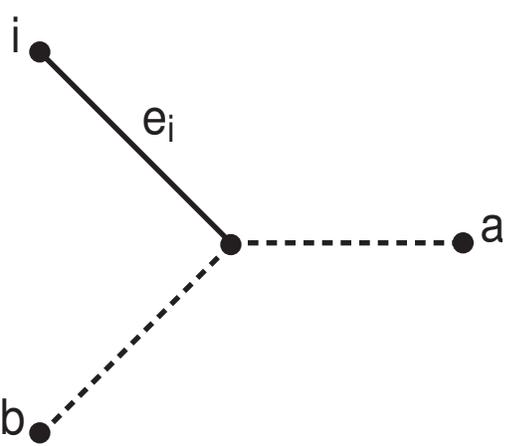
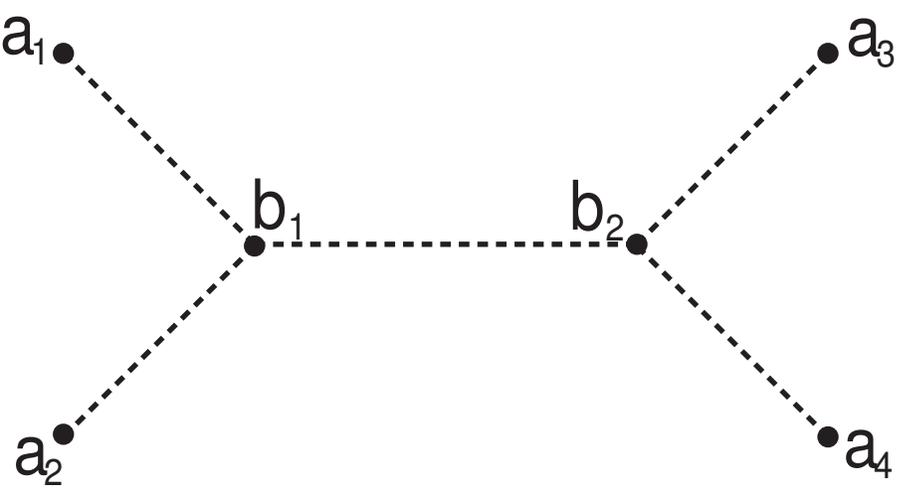
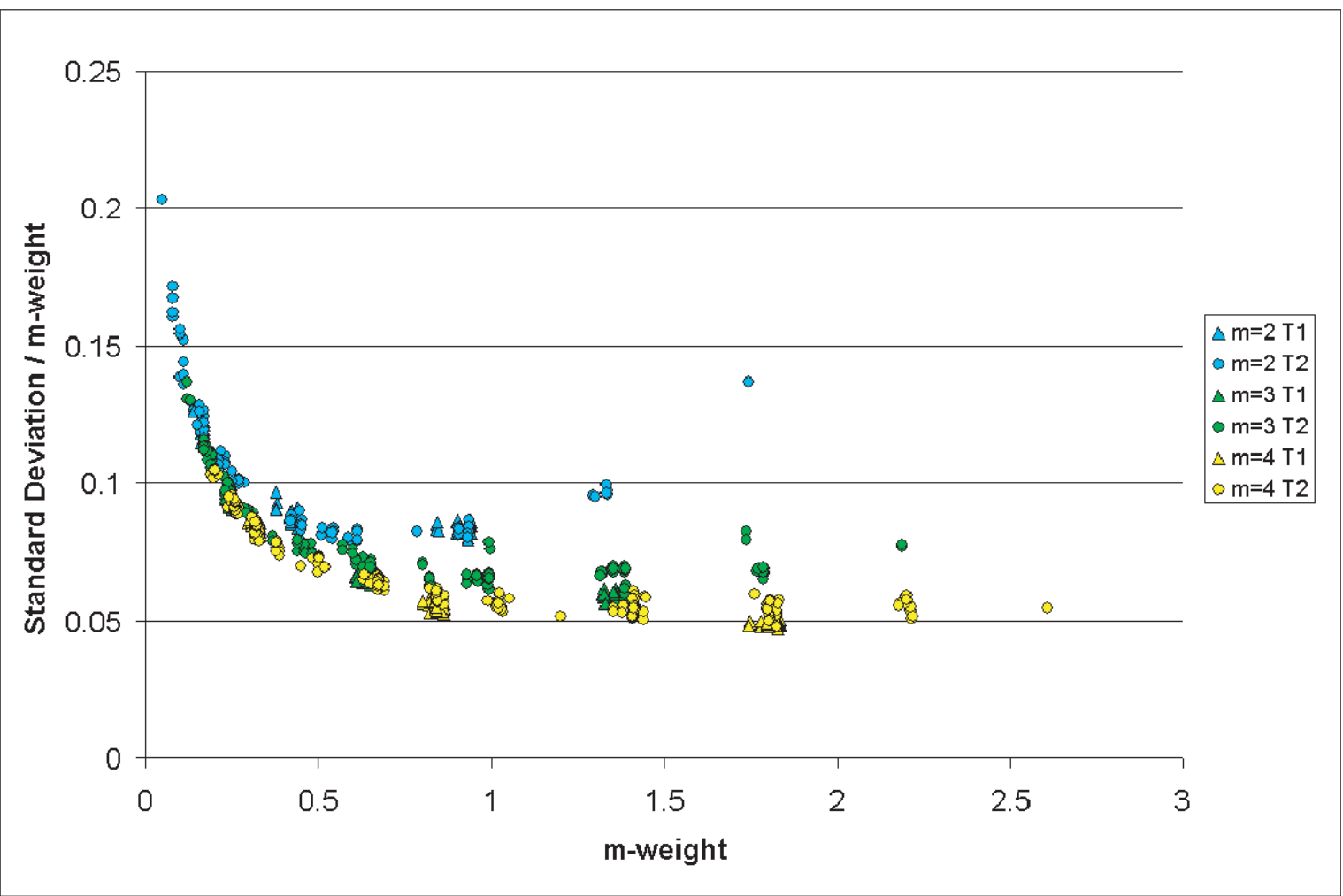
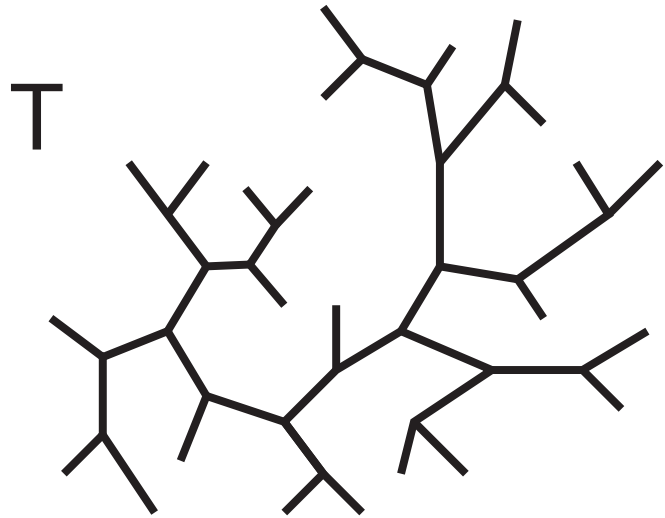
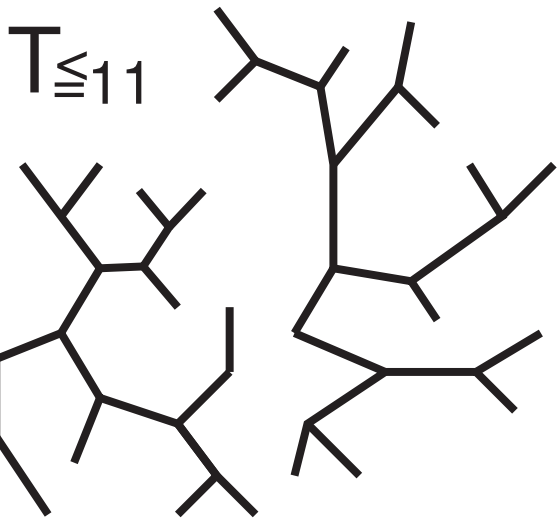
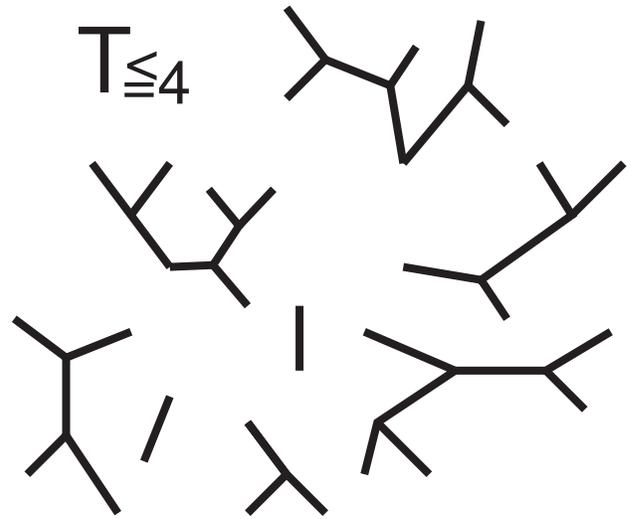
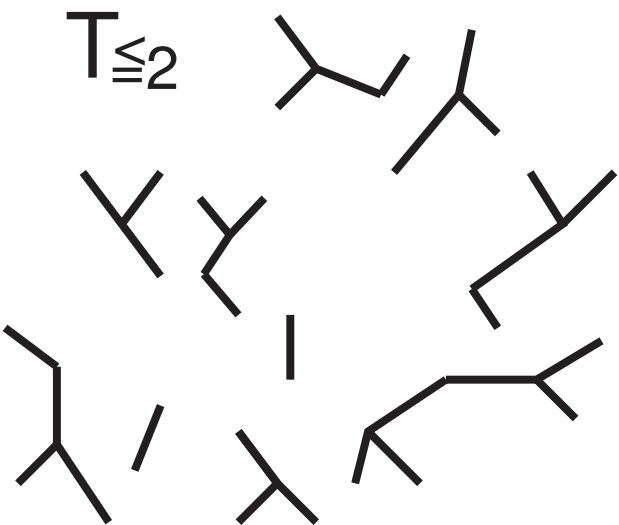
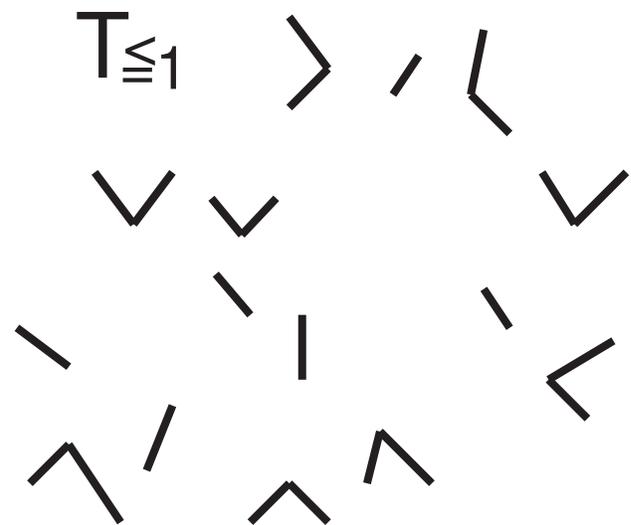
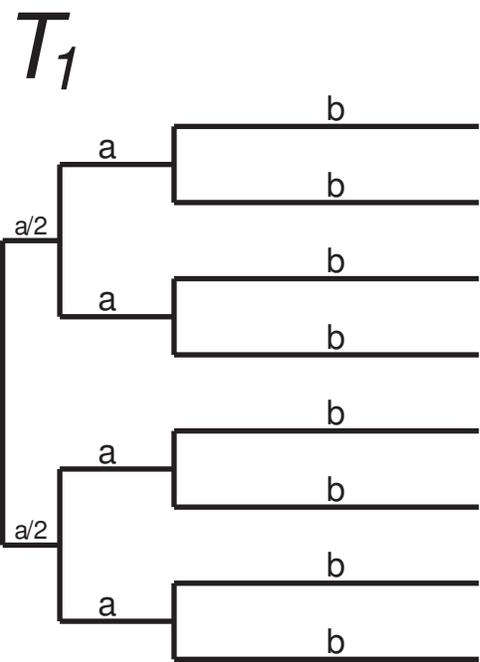
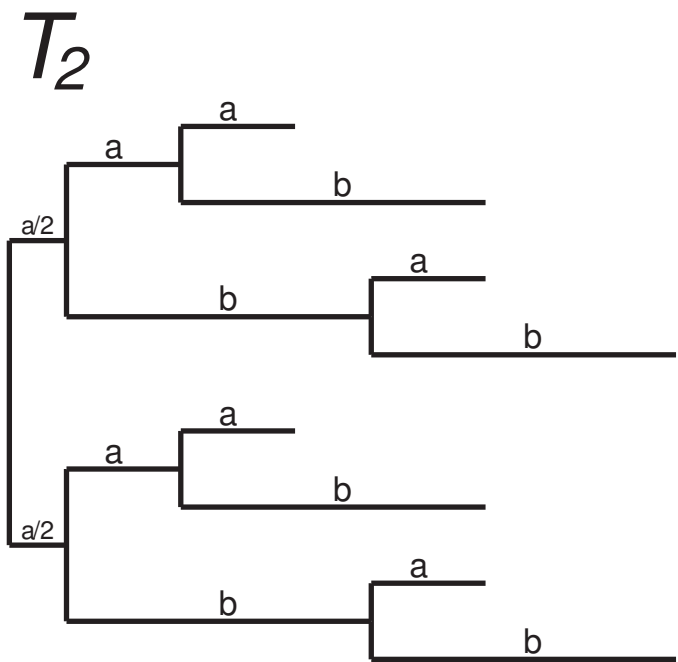
\end{document}